\documentclass[prd, amsfonts, twocolumn, nofootinbib, showpacs]{revtex4}
\usepackage{graphicx, epsfig}
\usepackage{caption}
\usepackage{color}
\usepackage{amsmath}
\usepackage{hyperref}
\newcommand{\be}{\begin{equation}}
\newcommand{\ee}{\end{equation}}
\newcommand{\bea}{\begin{eqnarray}}
\newcommand{\eea}{\end{eqnarray}}

\newcommand{\gapp}{\mathrel{\raise.3ex\hbox{$>$}\mkern-14mu \lower0.6ex\hbox{$\sim$}}}
\newcommand{\lapp}{\mathrel{\raise.3ex\hbox{$<$}\mkern-14mu \lower0.6ex\hbox{$\sim$}}}
\def\bbox{{\,\lower0.9pt\vbox{\hrule \hbox{\vrule height 0.2 cm
\hskip 0.2 cm \vrule  height 0.2 cm}\hrule}\,}}

\makeatletter
\newcommand*{\rom}[1]{\expandafter\@slowromancap\romannumeral #1@}
\makeatother

\usepackage{color}
\usepackage{ulem}
\usepackage[usenames,dvipsnames]{xcolor}
\usepackage{subfig}
\usepackage{tabularx}

\begin{document}
\title{Gravitational wave production by Hawking radiation from rotating primordial black holes}
\author{Ruifeng Dong, William H. Kinney, Dejan Stojkovic}
\affiliation{ HEPCOS, Department of Physics, SUNY, University at Buffalo, Buffalo, NY 14260-1500}
 %%%%%%%%%%%%%%%%%%%%%%%%%%%%%%%%%%%%%%%%%%%%%%%%%%%%%%%

\begin{abstract}
\widetext
In this paper we analyze in detail a rarely discussed question of gravity wave production from evaporating primordial black holes. These black holes emit
gravitons which are, at classical level, registered as gravity waves. We use the latest constraints on their abundance, and calculate
the power emitted in gravitons at the time of their evaporation. We then solve the coupled system of equations that gives us the evolution of the
frequency and amplitude of gravity waves during the expansion of the universe. The spectrum of gravitational waves that can be detected today depends
on multiple factors: fraction of the total energy density which was occupied by primordial black holes, the epoch in which they were formed, and
quantities like their mass and angular momentum. We conclude that very small primordial black holes which evaporate before the big-bang
nucleosynthesis emit gravitons whose spectral energy fraction today can be as large as $10^{-7.5}$. On the other hand, those which are
massive enough so that they still exist now can yield a signal as high as $10^{-6.5}$. However, typical frequencies of the gravity
waves from primordial black holes are still too high to be observed with the current and near future gravity wave observations.

\end{abstract}

%%%%%%%%%%%%%%%%%%%%%%%%%%%%%%%%%%%%%%%%%%%%%%%%%%

\maketitle

\section{Introduction}
Gravity waves are the last prediction of the general theory of relativity that has recently been verified by LIGO/Virgo collaborations \cite{Abbott:2016blz,Abbott:2016nmj}.
Many dynamical systems are capable of producing gravity waves, in particular mergers of black holes (BHs) or neutron
stars. It is also believed that primordial gravity waves can be produced in the early universe by tensor mode fluctuations during inflation, second-order
effects of scalar perturbations \cite{Bugaev:2010bb,Bugaev:2009kq}, or by phase transitions. Another mechanism that has not been extensively studied in the
literature is the BH evaporation. BH evaporation is close to thermal. This means that all the degrees of freedom whose mass is smaller
than the BH temperature will be democratically excited. Among the other degrees of freedom, gravitons will be excited too. These gravitons at the
classical level are nothing else but gravity waves. Therefore, primordial black holes (PBHs)  \cite{Carr:1974nx,Barrow:1992hq,Barrow:1996jk,Carr:1975qj,Khlopov:1985jw}
evaporation is one of the possible sources of gravity waves. This was studied for
the first time in \cite{Anantua:2008am}.

PBHs can form in the collapse of density fluctuations in the radiation-dominated universe (caused for example by some violent processes like phase transitions). 
This collapse is possibly anisotropic, producing rotating Kerr PBHs.  Even if the collapse is perfectly spherically symmetric and the resulting black 
hole is Schwarzschild, any subsequent accretion of the surrounding material will introduce non-zero angular momentum. Accretion is very efficient in 
spinning up the BHs.  The amount of accreted mass which is comparable to the initial BH mass is sufficient to spin up the BH 
from a non-rotating state to almost extremal Kerr state. This is especially important when we study PBH in a higher mass range (i.e. $> 10^{15}$ g) 
since their lifetime is comparable to or larger than the life-time of the universe, so they have enough time to significantly spin up.  It is 
therefore instructive to study the effect of BH rotation to the spectrum of emitted gravitational waves. In this paper we will, for the first 
time, calculate the possible gravity wave signals from rotating PBHs.

The thermal black body spectrum of particles emitted  from BHs is modified by greybody factors, which quantify the probability that a created
particle will penetrate the potential barrier and reach a distant asymptotic observer. Greybody factors can in certain cases modify the spectrum
very significantly. For example, it is known that a non-rotating Schwarzschild BHs preferably emit particles of lower spin (e.g. scalars), while
emission of particles with higher spin (e.g. gravitons) is suppressed. In contrast, a rotating Kerr BHs preferably emit gravitons. The reason is
that emission of higher spin modes from BHs gets amplified taking away rotational energy of BHs. This amplification enhances the probability of
graviton emission by many orders of magnitude. The typical energy of the particles (scalar, spinor, vector or tensor) is also larger in the emission
from faster rotating BHs. This enhanced emission in turn shortens the lifetime of BHs. Therefore, the greybody factor calculation is important to
get the gravity wave signals from evaporating Kerr PBHs.

In this paper, we extend the analysis performed in \cite{Anantua:2008am} in two ways. In order to study the effects of BH rotation, we first
calculate the greybody factors in section \ref{section:greybody}, and thus the gravity wave spectrum at emission time in section \ref{section:graviton}.
The evolution of the PBH parameters like mass and angular momentum is then also computable, as shown in section \ref{section:evolution}.
Second, we discuss the physical requirements in choosing parameters in section \ref{section:choosing}. The effects of cosmic expansion on the observed
signal is derived in section \ref{section:cosmic}. After completing this preliminary setup, we perform our calculations of gravity wave spectrum for
small PBHs that evaporate before the big-bang nucleosynthesis (BBN) in section \ref{section:evaporation}. In addition, after taking into account
the observational astrophysical limits (summarized in section \ref{section:constraints}) we also compute the gravity wave spectrum for larger black
holes that evaporated before or are still evaporating today in sections \ref{section:completely}.
The results are summarized in section \ref{conclusions}. Throughout the paper we use Planck units, i.e. $\hbar=c=G=k_B=1$, unless otherwise specified.

\section{Greybody factors for tensor modes}
\label{section:greybody}
Greyboy factors are essential in our analysis for two reasons. First, being the transmission probability over the gravitational potential barrier, they are
directly related to the graviton spectrum, as will be shown explicitly in section \ref{section:graviton}. Second, BHs evolve through evaporation
by emitting particles of all kinds, i.e. scalars, spinors, vectors and tensors.
Unfortunately, \cite{Teukolsky:1974yv} only gave partial results for tensor emission spectrum, and \cite{Page:1976ki} calculated the BH's emission into only
spinor, vector and tensor modes. The scalar case was considered much later by \cite{Taylor:1998dk}. Therefore, we will formulate the computation for
tensors in this section, and then display our results for the graviton spectrum in the following section.

Throughout the paper, we will consider a general Kerr BH, with the gravitational radius  $r_+$ (which is the outer event horizon size), mass $M$,
angular momentum $J$, and no electric charge. The dimensionless angular momentum $a_*\equiv J/M^2$ is also frequently used. In the Boyer-Lindquist coordinates
$(t, r, \theta, \varphi)$, the tensor field $\psi$ with energy $\omega$ and magnetic quantum number $m$ can be decomposed as
\be
\psi=e^{-i\omega t+i m\varphi}S(\theta)R(r),
\ee
The Teukolsky master equations \cite{Teukolsky:1973ha} for this field in the background of the Kerr BH are
\small
\be
\Delta^{2}\dfrac{d}{dr}\left(\frac1\Delta\dfrac{dR}{dr}\right)+\left(\dfrac{K^2+4 i(r - M) K}{\Delta}-8 i \omega r-\lambda \right)R=0,\label{ode:R}
\ee
\bea
\frac1{\sin\theta}\frac{d}{d\theta}\left(\sin\theta \frac{dS}{d\theta} \right)+&&\left[(2+a_* M\omega\cos\theta)^2\right.\nonumber\\
&&\left.-(m\csc\theta-2\cot\theta)^2-4+E \right]S=0,\nonumber\\ \label{ode:S}
\eea
\normalsize
where $\Delta\equiv r^2-2Mr+a_*^2 M^2$, $K\equiv (r^2+a_*^2 M^2)\omega-a_* M m$, $\lambda\equiv E+a_*^2 M^2 \omega^2-2 a_* M m\omega-2$, and
$E={_2}E^m_l(a_* M\omega)$ is to be determined from the eigenvalues of the function $S(\theta)$. $l$ is the angular momentum quantum number, no smaller
than $\max(|m|,|s|)$.

The above angular equation (Eq. (\ref{ode:S})) is the well-known spin-weighted spheroidal wave equation (spin-2). References \cite{Fackerell:1977,Press:1973zz} used Jacobian polynomial expansion to calculate the eigenvalues $_{s}E^m_l$ to the 6th order of $\gamma\equiv a_* M\omega$, for any spin $s$. This gives us a precision better than $\frac1{10^4}$ for $M\omega\le 3$. The results are summarized in App. \ref{appendix:spin}. However, nontrivial calculation is needed for the radial equation (Eq. (\ref{ode:R})),
which does not have analytical solutions, so numerical analysis is needed. We are interested in the greybody factors, which give the transmission
probability seen by an asymptotic observer if we consider a purely outgoing flux at infinity. Equivalently, by time reversal symmetry, this is the
absorption probability by the BH if we consider a purely ingoing flux at its horizon \cite{Dong:2015qpa}. As usual, we adopt the latter interpretation
in the following calculations.

The boundary conditions appropriate for our study are then
\bea
&&R\sim \Delta^2 e^{-i k r^*}, ~~~~~~~~~~~~~~~~~~~~~~~~r\rightarrow r_+,\label{nearhorizon}\\
&&R\sim Z_{in} r^{-1} e^{-i\omega r^*} + Z_{out} r^3 e^{i\omega r^*}, r\rightarrow \infty, \label{nearinfinity}
\eea
where $k\equiv\omega-m\Omega$, $\Omega\equiv a_*/(2r_+)$ is the angular velocity of the horizon, and $r^*$ is defined by
$\frac{dr^*}{dr}=\frac{r^2+a_*^2 M^2}{\Delta}$. $Z_{in}$ and $Z_{out}$ are constants to be found from numerical integration of Eq. (\ref{ode:R}) once we
fix the boundary solution near the horizon as (\ref{nearhorizon}). At the two boundaries, the field energy fluxes are \cite{Teukolsky:1974yv}
\bea
\frac{d^2 E_{hor}}{dtd\Omega_3}&=&\frac{{_2}S^2_{lm}(\theta)}{2\pi} \frac{128\omega k(k^2+4\epsilon^2)(k^2+16\epsilon^2)(2Mr_+)^5}{|C|^2}, \nonumber\\&&~~~~~~~~~~~~~~~~~~~~~~~~~~~~~~~~~~~~~~~~r\rightarrow r_+\\
\frac{d^2 E_{out}}{dtd\Omega_3}&=&\frac{{_2}S^2_{lm}(\theta)}{2\pi}\frac1{2\omega^2}|Z_{out}|^2, ~~~~~~~~~~~~~~r\rightarrow\infty\\
\frac{d^2 E_{in}}{dtd\Omega_3}&=&\frac{{_2}S^2_{lm}(\theta)}{2\pi}\frac{128\omega^6}{|C|^2}|Z_{in}|^2, ~~~~~~~~~~~~r\rightarrow\infty,
\eea
where $\epsilon\equiv\sqrt{1-a_*^2}/(4r_+)$, and
\small
\bea
|C|^2&=&(Q^2+4a_*M\omega m-4a_*^2M^2\omega^2)\left[(Q-2)^2+36a_*M\omega m \right.\nonumber\\
&-&\left. 36a_*^2M^2\omega^2\right]+(2Q-1)(96a_*^2M^2\omega^2-48a_*M\omega m)\nonumber\\
&+& 114M^2\omega^2(1-a_*^2),
\eea
\normalsize
with $Q\equiv {_2}E^m_l+a_*^2M^2\omega^2-2a_*M\omega m$. Here $\Omega_3$ is the solid angle in $3$-dimensional space and $_{2}S_{lm}(\theta)$ is the
normalized eigenfunction of Eq. (\ref{ode:S}), which does not concern us here. The field absorption probability is
\small
\bea
&\gamma_{2lm}&(a_*,M\omega) \equiv 1-{\frac{d^2 E_{out}}{dtd\Omega_3}}\bigg/{\frac{d^2 E_{in}}{dtd\Omega_3}} \nonumber\\
 &=& \frac{d^2 E_{hor}}{dtd\Omega_3}\bigg/\left(\frac{d^2 E_{hor}}{dtd\Omega_3}+\frac{d^2 E_{out}}{dtd\Omega_3}\right) \nonumber\\
 &=& \frac{128\omega k(k^2+4\epsilon^2)(k^2+16\epsilon^2)(2Mr_+)^5}{128\omega k(k^2+4\epsilon^2)(k^2+16\epsilon^2)(2Mr_+)^5 + \frac{|C|^2}{2\omega^2}|Z_{out}|^2}.\nonumber\\
\label{gamma2}
\eea
\normalsize

In the numerical calculation, we set our scale by letting $r_+ =1$. After imposing the solution (\ref{nearhorizon}) at $r=1+10^{-3}$, we integrate
(\ref{ode:R}) to larger $r$, until the solution becomes stable, in practice at $r=10^3$. The $r^3$ term dominates $R(r)$ there, so the factor $Z_{out}$
is easily obtained. The greybody factor is then computed using Eq. (\ref{gamma2}) for any mode $(\omega, l, m)$.

\section{Instantaneous graviton spectrum}
\label{section:graviton}
Two factors are affecting the graviton spectra that we can observe today. One is the Hawking spectrum of a PBH with fixed mass $M$ and angular
momentum $J$, and the other is the evolution of $M$ and $J$ in the expanding universe. Here we focus on the former, and the latter is dealt with in
the next section.

We first define a dimensionless quantity $Q^{GW}(a_*, M\omega)$, which represents the expected emission rate, i.e. the number of particles emitted per
unit frequency per unit time, for gravitons. This quantity includes blackbody and greybody terms which are uniquely determined with two dimensionless
parameters $a_*\le 1$ and $x\equiv M\omega$. Formally,
\small
\bea
&Q^{GW}&(a_*,x)\equiv\frac{d^2 N}{dtd\omega}\Bigr |_{s=2}\nonumber\\
&=\sum\limits_{plm}&\frac{\gamma_{2lm}(a_*,M\omega)}{\exp((\omega-m\Omega)/T_H)-1}  \nonumber\\
&=\sum\limits_{plm}&\frac{\gamma_{2lm}(a_*,x)}{\exp\left(4\pi\left[1+(1-a_*^2)^{-\frac12}\right]x-2\pi m a_* (1-a_*^2)^{-\frac12}\right)-1},\nonumber\\
\eea
\normalsize
where $T_H=\dfrac1{2\pi}\left(\dfrac{r_+-M}{r_+^2+a_*^2 M^2}\right)$ is the Hawking temperature. For gravitons, the number of polarizations $p$ is $2$.

Using the program described in the previous section, we calculated $Q^{GW}(a_*, M\omega)$ for 156 values of $a_*$ from 0 to 0.99999, and 3000 values of $x$
from 0.001 to 3.\footnote{The eigenvalues of spin-weighted spheroidal equation are calculated with a precision better than $\frac1{10^4}$ for $a_* x$
only up to 3, using the polynomial expansion to the 6th order.\ref{appendix:spin}} For low $a_*$, only the several low-$l$ modes are important. For each $l$, modes with
different $m$'s make similar contributions, as spherical symmetry of the background is not severely distorted. For high $a_*$, angular momentum can
be transferred from the BH to $m\approx l$ modes, and high-$l$ modes become more important. For example, to keep $Q^{GW}(a_*, x)$ at a precision
better than $\frac1{10^5}$, we considered modes with $l=2, 3, 4, |m|\le l$ for $a_*=0$, and modes with $m=l=2, 3, ..., 14$ for $a_*=0.99999$. Our
results for the greybody factors are consistent with \cite{Teukolsky:1974yv}, and results for the total energy and angular momentum emission rates are consistent
with \cite{Page:1976ki}.

\begin{figure}[h]
\captionsetup{justification=raggedright,
singlelinecheck=false
}
\includegraphics[width=3.4in]{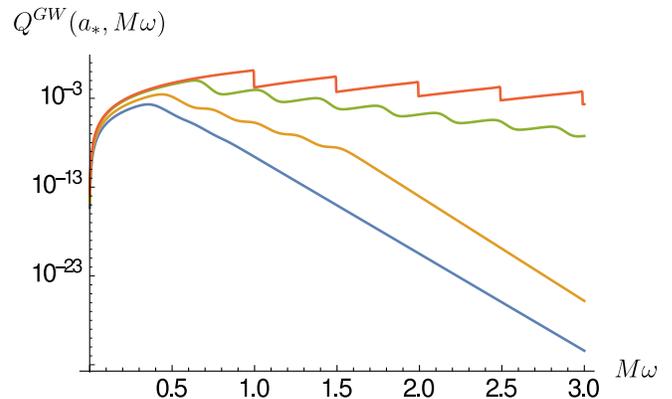}
\caption{Expected graviton emission rate vs. particle energy. From bottom to up, the curves correspond to $a_*=0, 0.5, 0.9, 0.99999$ respectively. For
$a_*=0.99999$, five spikes from left to right originate from emission of $m=l=2, 3, 4, 5, 6$ modes, respectively.}
\label{figQ-a-x}
\end{figure}

Fig. (\ref{figQ-a-x}) shows the graviton emission rates for $a_*=0, 0.5, 0.9, 0.99999$. For $a_*$ closer to unity, superradiance causes enhanced emission and
large particle energy range. In particular, each spike in $Q^{GW}(0.99999, x)$ corresponds to emission of one $m=l$ mode. Clearly, faster rotating
BHs tend to emit more gravitons with higher energies. This is also the case for emission of other particles (though to a lesser extent), due to coupling between the spin of the
particle and that of the hole. This happens for both bosons and fermions, even though superradiance happens only for boson fields, as explained in \cite{Dai:2010xp}.
As a result, spin enhances the BH's total emission rate and thus reduces its lifetime, which will be seen in more detail in the next section.

\begin{figure}[h]
\captionsetup{justification=raggedright,
singlelinecheck=false
}
\includegraphics[width=3.4in]{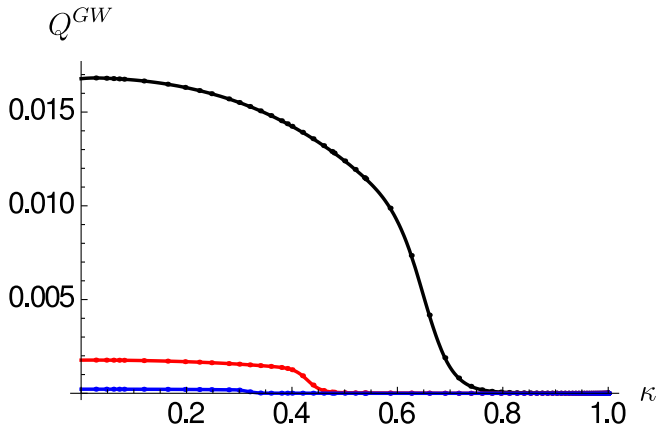}
\caption{Interpolated function $Q^{GW}$ of $\kappa=\frac2{1+(1-a_*^2)^{-1/2}}$, for different $x$'s. From bottom to top, three sets of data points and
smooth curves correspond to $x=1, 2, 3$, respectively.}
\label{figQ-kappa}
\end{figure}

For each value of $x$, we used linear interpolation in $x$ and cubic spline interpolation in $\kappa\equiv \frac2{1+(1-a_*^2)^{-1/2}}$ to get a smooth
function $Q^{GW}(a_*, x)$. Fig. (\ref{figQ-kappa}) shows this function for $x=1, 2, 3$. This function will be used later in calculating today's graviton
spectrum.

\section{Evolution of Kerr BHs}
\label{section:evolution}
Through Hawking radiation, BH's mass and angular momentum evolve with time. We use the dimensionless parameters $f$ and $g$ defined by Page
\cite{Page:1976ki} to trace the evolution of Kerr BHs,
\small
\bea\label{def_f}
&f&\equiv -M^2 \frac{dM}{dt}\nonumber\\
&&=\sum\limits_{splm}\int_0^{\infty} \frac{d\omega}{2\pi} \frac{\omega M^2\gamma_{slm}(a_*,M\omega)}{\exp((\omega-m\Omega)/T_H)\pm 1} \nonumber\\
&&=\sum\limits_{splm}\int_0^{\infty} \frac{dx}{2\pi}\times x\gamma_{slm}(a_*,x) \nonumber\\
&\times&\left[\exp\left(4\pi\left[1+(1-a_*^2)^{-\frac12}\right]x-2\pi m a_* (1-a_*^2)^{-\frac12}\right)\pm 1\right]^{-1} ,\label{f_a}\nonumber\\
\eea
\bea\label{def_g}
&g&\equiv -\frac{M}{a_*}\frac{dJ}{dt}\nonumber\\
&&=\sum\limits_{splm}\int_0^{\infty} \frac{d\omega}{2\pi} \frac{mMa_*^{-1}\gamma_{slm}(a_*,M\omega)}{\exp((\omega-m\Omega)/T_H)\pm 1}\nonumber \\
&&=\sum\limits_{splm}\int_0^{\infty}\frac{dx}{2\pi}\times ma_*^{-1}\gamma_{slm}(a_*,x)\nonumber\\
&\times&\left[\exp\left(4\pi\left[1+(1-a_*^2)^{-\frac12}\right]x-2\pi m a_* (1-a_*^2)^{-\frac12}\right)\pm 1\right]^{-1}. \label{g_a}\nonumber\\
\eea
\normalsize
Here the $\pm$ signs account for different statistics of bosons and fermions. Functions $f$ and $g$ are measuring the rate of change of mass and
angular momentum respectively. They depend on $a_*$ only, as the greybody factors only depend on $a_*$ and $x=M\omega$. We calculated the contributions
to $f$ and $g$ from $s=0, 2$, consistent with \cite{Taylor:1998dk} and \cite{Page:1976ki} respectively. For $s=1/2, 1$, we just used the results in
\cite{Page:1976ki}. Tab. (\ref{tabf-g}) lists all relevant $f$ and $g$ contributions.

\begin{figure}[h]
\captionsetup{justification=raggedright,
singlelinecheck=false
}
\includegraphics[width=3.4in]{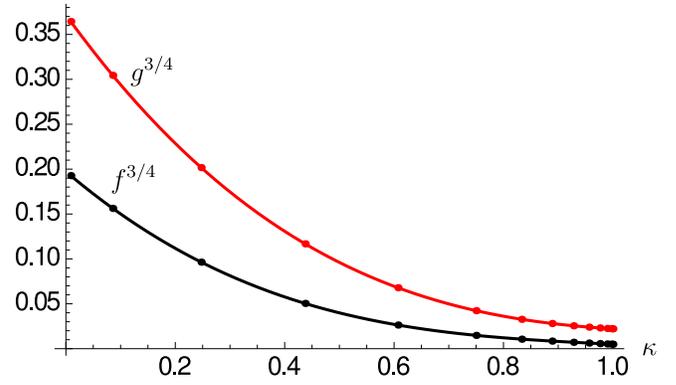}
\caption{The first fourteen values of $f^{3/4}$ and $g^{3/4}$ from Tab. (\ref{tabf-g}) are fitted as third order polynomials in $\kappa=\frac2{1+(1-a_*^2)^{-1/2}}$.
The fitted parameters all have uncertainties below $\frac1{300}$.}
\label{figf-g-kappa}
\end{figure}

Practically, we consider the number of degrees of freedom, i.e. the number of different particle species and polarizations, for scalars, spinors,
vectors and tensors to be 1, 18, 2, 2 respectively as in the standard model\footnote{For simplicity, we take one Higgs boson for scalars, all leptons
for spinors, photons for vectors, and gravitons for tensors.}. Then $f^{3/4}$ and $g^{3/4}$ are fitted as function of $\kappa$ using 3rd order
polynomial, with relative uncertainty below $\frac1{300}$ for all fitted parameters. The result is plotted in Fig. (\ref{figf-g-kappa}).

Once functions $f(a_*)$ and $g(a_*)$ are known, we can solve the evolution of BHs, from the definitions of $f$, $g$, i.e. Eqs. (\ref{def_f}, \ref{def_g}),
which can be rewritten as
\bea
\frac{d\ln\bar{M}}{d\ln a_*}=\frac{f(a_*)}{g(a_*)-2f(a_*)}\nonumber\\
\frac{d\tau}{d\ln a_*}=\frac{\bar{M}^3}{2f(a_*)-g(a_*)},\label{M-a-t}
\eea
where $\bar{M}\equiv M/M_i$ and $\tau\equiv t/M_i^3$. With initial conditions $\tau_i=0$, $a_{*,i}=0.99999$ and $\bar{M}_i=1$, we solve Eqs.
(\ref{M-a-t}) for $a_*$ down to 0.004. Thereafter, $f$, $g$ are almost constant and we can use the analytic solution
\bea\label{M-t-anal}
\bar{M}(\tau)&=&(const_1-3f(0)\tau)^{1/3}\nonumber\\
a_*(\tau)&=&const_2\times \left(\bar{M}(\tau)\right)^{\frac{g(0)-2f(0)}{f(0)}},
\eea
where $const_1, const_2$ are fixed by connecting with the previous numerical solution. There, an interval of 0.005 in $\ln a_*$ gives us a precision
better than $\frac1{10^4}$.

\begin{figure}[h]
\captionsetup{justification=raggedright,
singlelinecheck=false
}
\includegraphics[width=3.4in]{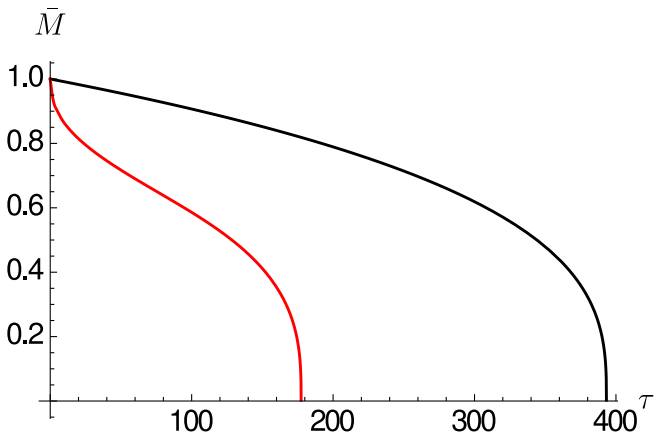}
\caption{The BH's mass evolution with time. The left curve is for $a_{*,i}=0.99999$ and the right one for $a_{*,i}=0$. In the begining, the rotating BH loses
mass more efficiently, due to its angular momentum transfer to particles. Near the end of evaporation, the BH has shed its spin, and so evolves in the
same way as a Schwarzschild BH. This is why the two curves are nearly parallel as $\bar{M}\rightarrow 0$.}
\label{figM-tau}
\end{figure}

Ten percent of the numerically solved data points are interpolated using cubic splines to form a smooth function $\bar{M}(\tau)$, which is shown in Fig.
(\ref{figM-tau}). Also shown is the BH evolution if $a_{*,i}=0$, where Eqs. (\ref{M-t-anal}) are the solutions. For non-rotating BHs, the mass decay gets
faster with time, because the Hawking temperature gets higher as the mass decreases. For Kerr BHs, on the other hand, the initial energy emission
rate is high due to angular momentum transfer. Later, after the BH sheds its spin, the BH follows the history of Schwarzschild BHs.

Apparently, the BH's lifetime, denoted as $\tau_e$, is shortened by its initial spin. For comparison, $\tau_e=393.3$ for $a_{*,i}=0$, while
$\tau_e=177.4$ for $a_{*,i}=0.99999$.

\section{Choosing parameters for PBH production}
\label{section:choosing}
Two observational constraints are important in this context. First, the initial energy density in the universe should be less than the energy scale
at the end of inflation, i.e. $\rho_i\lesssim\left(10^{16}GeV\right)^4$, due to non-detection of gravitational waves at that scale. Second, in order
not to contradict cosmological
observations, the safest assumption is that PBHs disappear before BBN \cite{Anantua:2008am} (we will relax this requirement later).
We choose the temperature at BBN as when neutrinos began to decouple from the electrons, i.e. $10^{10}K$. The total energy density at BBN is
\cite{Weinberg2008}
\be
\rho_{BBN}=a_B T^4+\frac78\times 3 a_B \left(\frac{T}{1.008} \right)^4=(0.753 MeV)^4,
\ee
\\
where $a_B$ is the Stefan-Boltzmann constant and $\frac{T}{1.008}$ is the temperature of neutrinos at that time. Therefore, we require that the total
energy density at the end of PBH's evaporation, $\rho_e$, is higher than $(0.753 MeV)^4$.

In the radiation-dominated universe, PBHs form when a sufficiently large density fluctuation falls into the particle horizon, so
we take $M_i\leq\frac{4\pi}3\rho_i H_i^{-3}$, where the sign $<$ comes from the possibility that early phase transition can greatly enhance the
formation of smaller PBHs \cite{Niemeyer:1997mt}. For simplicity, the equal sign is taken in the following calculations.
The initial collapse of density fluctuations might not be completely isotropic. As a result, the formed PBHs can possibly have non-zero angular momentum.
In addition, accretion is very effective in spinning BHs up. The amount of accreted mass which is comparable to the initial BH mass is sufficient to 
spin up the BH from a non-rotating state to almost extremal Kerr state \cite{Ab}.
In the following, we consider PBHs in two extreme cases, i.e. $a_{*,i}=0$, $0.99999$, to illustrate the effect of the BH spin.

In Sec. \ref{section:greybody} we calculated the absorption
probability (greybody factor) of BHs considering incoming waves at flat asymptotic infinity, which is satisfied only if $r_+ \ll d_{BH}$, where $d_{BH}$ is the
average physical distance between neighboring BHs. So we need $M_i\ll(\beta\rho_i/M_i)^{-1/3}$. Here $\beta$ is the PBH mass fraction at formation.
This is easily satisfied if $\beta\ll 1$. This is the third constraints that we will impose.

In the following, we discuss both PBHs which evaporated before BBN, and those which evaporated by today or still exist. For the former,
we take $\beta= 10^{-2}, 10^{-4}, 10^{-8}, 10^{-16}$. For the latter, we take the upper bounds of observational constraints on $\beta$ as a function
of $M_i$, in order to get the largest possible signal of gravity waves. For simplicity, we will always assume a uniform initial mass distribution for
PBHs forming at one certain epoch.

\section{Cosmic expansion}
\label{section:cosmic}
The PBHs are evaporating in the expanding cosmos. Now we complete the link between the present-day observable spectra and the instantaneous emission
rate $Q^{GW}(a_*,x)$.

From the graviton energy and density, we denote the present-day values with a tilde on the top, and their instantaneous values without. Consider a time interval $dt$ and
frequency interval $d\omega$, in which some amount of energy $d\omega dt \frac{dE}{dt d\omega}$ is emitted in gravitons. We have
\be
\frac{d\rho_{GW}}{dt d\omega}= n_{BH}\frac{dE}{dt d\omega},
\ee
where $n_{BH}$ is the instantaneous number density of PBHs. This energy density of massless gravitons scales as $a^{-4}$ as the universe
expands, where $a$ is the cosmological scale factor. We can then scale the graviton energy density to today $\tilde{\rho}_{GW}=\rho_{GW}a^4$.\footnote{The
cosmological scale factor today is taken to be $1$.} And its frequency $\omega$ is related to its redshifted today's value $\tilde{\omega}$ as
$\tilde{\omega}=a\omega$. We ignore anisotropic stress from neutrino free streaming \cite{Weinberg:2003ur}, since the wavelengths of the modes are much
smaller than the Hubble length. We immediately get
\bea
dt \frac{d\tilde{\rho}_{GW}}{dt d\tilde{\omega}}&=&dt a^3(t) n_{BH}(t)\frac{dE}{dt d\omega} \nonumber\\
&=& dt a_i^3 n_{BH,i} \frac{dE}{dt d\omega} \nonumber\\
&=& \frac{\beta\rho_i a_i^3}{M_i} dt \frac{dE}{dt d\omega},
\eea
where the second equality follows from $n_{BH}\sim a^{-3}$. The instantaneous graviton spectrum is related to the $Q$ factor as
\bea
\frac{dE}{dt d\omega}&=&\frac{\omega}{2\pi}Q^{GW}(a_*(t),M(t)\omega) \nonumber\\
&=& \frac{\tilde{\omega}}{2\pi a(t)}Q^{GW}\left(a_*(t),M_i\tilde{\omega}\frac{\bar{M}(t)}{a(t)}\right).
\eea
Here $\bar{M}(t)$ and $a_*(t)$ have already been solved in Sec. (\ref{section:evolution}). Integrating over $t$, we get
\bea\label{Omega-GW}
\Omega_{GW}&\equiv&\frac{\tilde{\omega}}{\rho_c}\frac{d\tilde{\rho}_{GW}}{d\tilde{\omega}} \nonumber\\
&=&\frac{\beta a_i^3 \tilde{\omega}^2}{8\pi H_0^2} \int^{\tau_f}_0 \frac{d\tau}{a(\tau)} Q^{GW}\left(a_*(\tau),M_i\tilde{\omega}\frac{\bar{M}(\tau)}{a(\tau)}\right),\nonumber\\
\eea
where $\rho_c$ is the critical density of the universe, which is related to today's Hubble parameter $H_0$ by the Friedmann equation.
$H_0=h\times 100$ km/s/Mpc. $h$ is left as a free parameter for PBHs evaporating before BBN, but it will be taken as the observed value of 0.673 for PBHs
which completed evaporation only recently or are still evaporating today. The upper limit
of integration $\tau_f$ is the PBH's lifetime $\tau_e$ if they've evaporated by today, or today's time otherwise.\footnote{Strictly, the integration lower limit
is fixed by the total energy density at PBH's formation, i.e. $t_i=1/(2H_i)=(32\pi\rho_i/3)^{-1/2}$. But $\tau_i=t_i/M_i^3$ is always very close to 0.}
In order to calculate today's gravity wave spectrum $\Omega_{GW}$, we need to know $a(\tau)$, which can be found from the Friedmann equation
coupled with equations governing the evolution of the energy densities of PBHs and radiation, i.e.
\bea
\frac{d\rho_{BH}}{dt}&=&-3\frac{\dot{a}}{a}\rho_{BH}+\rho_{BH}\frac{\dot{M}}{M}, \nonumber\\
\frac{d\rho_{rad}}{dt}&=&-4\frac{\dot{a}}{a}\rho_{rad}-\rho_{BH}\frac{\dot{M}}{M}, \nonumber\\
\frac{\dot{a}}{a}&=&\sqrt{\frac{8\pi}{3}(\rho_{BH}+\rho_{rad})}.
\label{complete-evolution}
\eea
Here dot overhead means the derivative with respect to cosmic time $t$. We will numerically solve Eqs. (\ref{complete-evolution}) in the following.\footnote{With
certain simplified special assumptions on the form of PBH mass spectrum and of greybody factors, the above coupled equations can be analytically solved \cite{Barrow:1991dn}.}

\section{Evaporation before BBN (PBHs with $M_i\lesssim 10^9 g$)}
\label{section:evaporation}
The process of PBH formation is not fully understood, so we won't constrain ourselves to specific formation epoches (e.g. at early
universe phase transitions \cite{Rubin:2000dq,Khlopov:1999ys,Khlopov:2008qy,Carr:2005zd,Jedamzik:1996mr,Stojkovic:2005zh,Stojkovic:2004hz,Bugaev:2013fya,Bugaev:2011wy}).
Instead, in this section we will only require that PBHs evaporate before BBN, after which the universe is radiation-dominated
until matter-radiation equality. Therefore, for a given initial total energy density and PBH fraction, the production epoch ($a_i$) of PBHs will be fixed
by the energy density right after evaporation, that is, $\rho_{e}a_{e}^4=\rho_{BBN}a_{BBN}^4$.

The cosmological evolution before BBN is not well constrained, so we have freedom to choose the initial energy fraction of PBHs, not
worrying that they can take over the dominance at some time before complete evaporation\footnote{PBH mergers can largely increase their lifetime,
easily making them survive after BBN. But the merger rate goes as $\beta^{N-1}$, where N is the number of PBHs which are merged, as
estimated in \cite{Anantua:2008am}. We have chosen $\beta$ to be no larger than $10^{-2}$, so this effect can be safely neglected.}. In the following,
we shall choose it to be $10^{-2}$, $10^{-4}$, $10^{-8}$, $10^{-16}$. Also, we are interested in PBHs produced at various scales, so we shall choose several
values for the initial total energy density, that is $\rho_i=10^{-12}, 10^{-20}, 10^{-28}$. The unit of $\rho_i$ is $M^4_P$, where $M_P$ is the Planck mass.

Using 4th-order Runge-Kutta method, we solved Eqs. (\ref{complete-evolution}). For each pair of ($\rho_i, \beta$) and initial scale factor $a_i=1$, an
interval of 0.1 in $\ln\tau$ for $\tau\le 100$ and in $\tau$ for $\tau>100$ gives us a precision better than $\frac{1}{10^6}$. In order to calculate
the BH's signal observed at present, we need to know the initial scale factor relative to today's, $a_0$. Take $a_0=1$, then at BBN,
\be
a_{BBN}=\frac{T_{\nu 0}}{T/1.008}=1.96\times 10^{-10},
\ee
where $T_{\nu 0}=1.945 ~K$ is neutrino's temperature today.

If PBHs finished evaporation at some time $t_{e}$ before BBN, we assume there was no phase change in the universe between this time
and BBN. Thus we have $\rho_{rad,e}a_{e}^4=\rho_{BBN}a_{BBN}^4$. Therefore, after numerical integration, we get the scale factor
and radiation energy density at the end of evaporation (when $\rho_{BH,e}$ just vanished). We then determine the correct initial scale factor as
$a_i=\left(\dfrac{\rho_{BBN}a_{BBN}^4}{\rho_{rad,e}a_{e}^4}\right)^{1/4}$.

\begin{figure}
\captionsetup{justification=raggedright,
singlelinecheck=false
}
  \begin{tabular}{@{}c@{}}
    \includegraphics[width=3.4in]{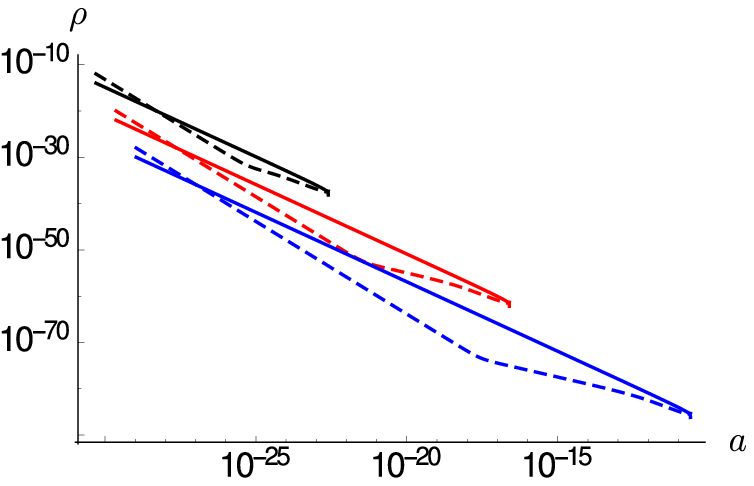}\\
    \small (a) $\beta=10^{-2}$
  \end{tabular}

  \begin{tabular}{@{}c@{}}
    \includegraphics[width=3.4in]{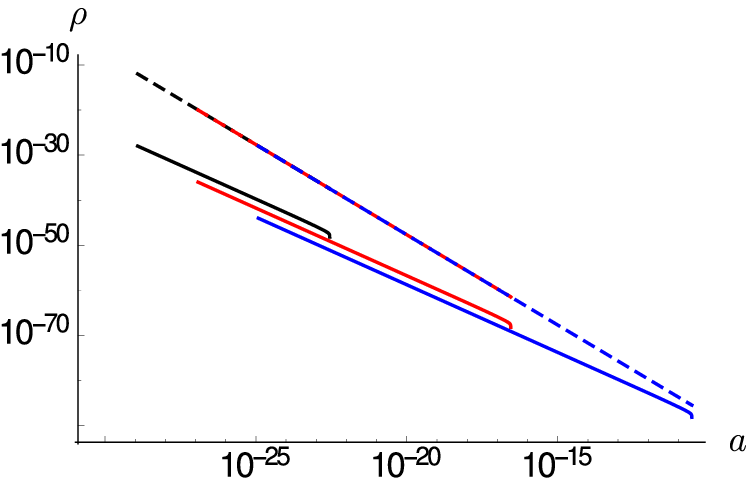}\\
    \small (b) $\beta=10^{-16}$
  \end{tabular}
  \caption{Evolution of PBH (solid curves) and radiation (dashed curves) energy densities with the cosmological scale factor $a$, during the
  lifetime of PBHs, for $a_*=0.99999$. The upper plot is for $\beta=10^{-2}$, and the lower for $\beta=10^{-16}$. Different colors denote different $\rho_i$'s. Black, red,
  blue cures are for $\rho_i=10^{-12}, 10^{-20}, 10^{-28}$ respectively.}
  \label{fig:rho-a}
\end{figure}

The results of the evolution of energy densities, are shown for $\beta=10^{-2}, 10^{-16}$ and $a_{*,i}=0.99999$ in Figs. (\ref{fig:rho-a}). With log-scales, the BH evaporation
is only seen as a dip at the end of the solid curves.\footnote{Also for this reason, the evolution for the case $a_{*,i}=0$ has no visible difference with the
$a_{*,i}=0.99999$ case, thus not shown here.} The cosmological scaling is clearly seen, i.e. $\rho_{BH}\sim a^{-3}$ and $\rho_{rad}\sim a^{-4}$.
 For $\beta=10^{-16}$, PBHs never dominate, so the radiation-dominated universe evolves just as in standard cosmology. For $\beta=10^{-2}$, however,
PBHs take over the dominance before completing evaporation. Therefore, universe starting with different energy scales evolves along different curves.

\begin{figure}[h]
\captionsetup{justification=raggedright,
singlelinecheck=false
}
\includegraphics[width=3.4in]{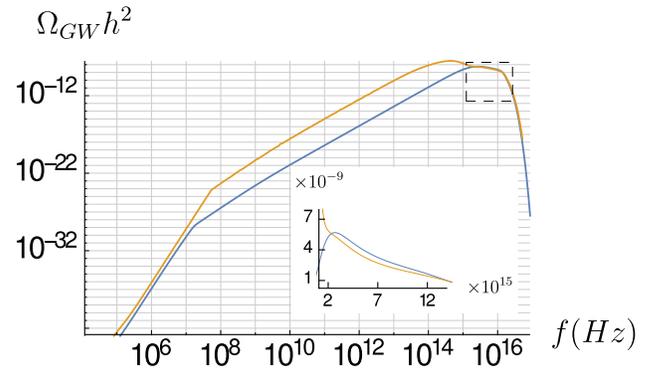}
\caption{Today's gravity wave spectrum, for $\rho_i=10^{-12}$, $\beta=10^{-2}$. Different colors denote different $a_{*,i}$'s. Blue and orange cures are
for $a_{*,i}=0$, $0.99999$ respectively. The part within a dashed square is also shown in a linear-scale sub-plot. Note the dashed square is not
the exact position of the sub-plot.}
\label{fig:BBN-Omega-f_a}
\end{figure}

\begin{figure}[h]
\captionsetup{justification=raggedright,
singlelinecheck=false
}
  \begin{tabular}{@{}c@{}}
    \includegraphics[width=3.4in]{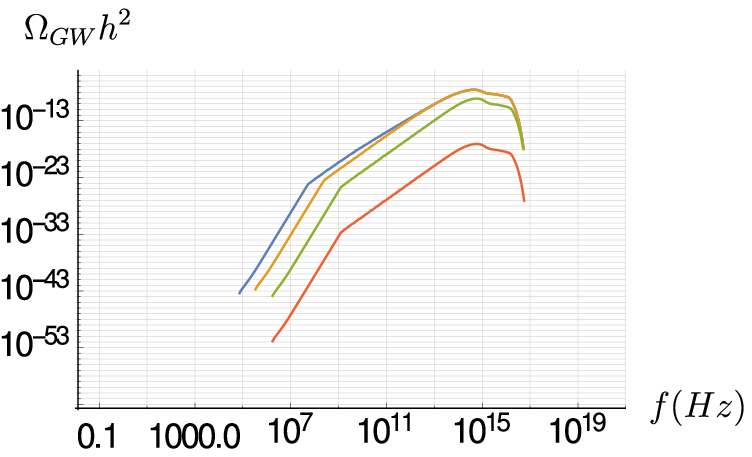}\\
    \small (a) $\rho_i=10^{-12}$
  \end{tabular}
  \begin{tabular}{@{}c@{}}
    \includegraphics[width=3.4in]{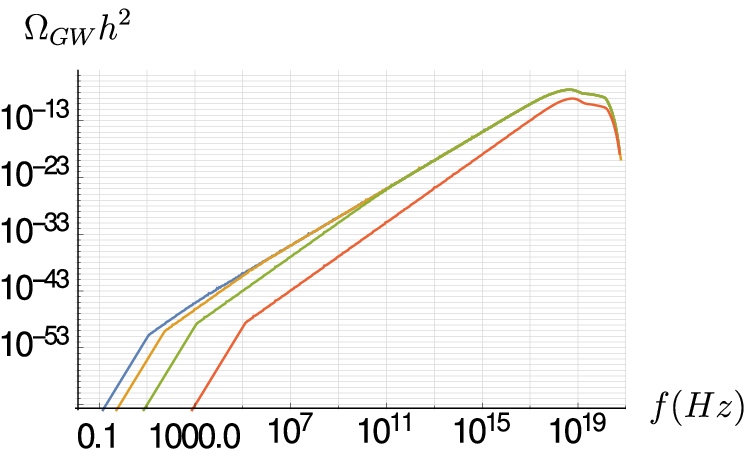}\\
    \small (b) $\rho_i=10^{-28}$
  \end{tabular}
  \caption{Today's gravity wave spectrum, for $a_{*,i}=0.99999$. $\rho_i=10^{-12}$ for the upper plot, and $\rho_i=10^{-28}$ for the lower one. In each
  plot, four curves from top to bottom, are for $\beta=10^{-2}, 10^{-4}, 10^{-8}, 10^{-16}$, respectively.}
\label{fig:BBN-Omega-f_beta}
\end{figure}

\begin{figure}[h]
\captionsetup{justification=raggedright,
singlelinecheck=false
}
  \begin{tabular}{@{}c@{}}
    \includegraphics[width=3.4in]{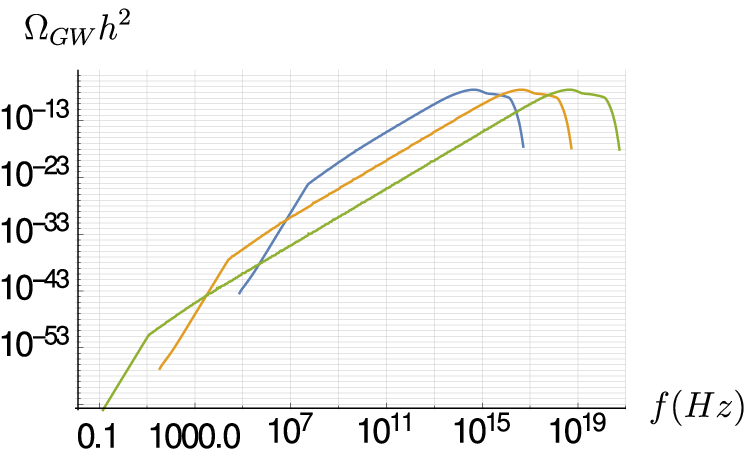}\\
    \small (a) $\beta=10^{-2}$
  \end{tabular}
  \begin{tabular}{@{}c@{}}
    \includegraphics[width=3.4in]{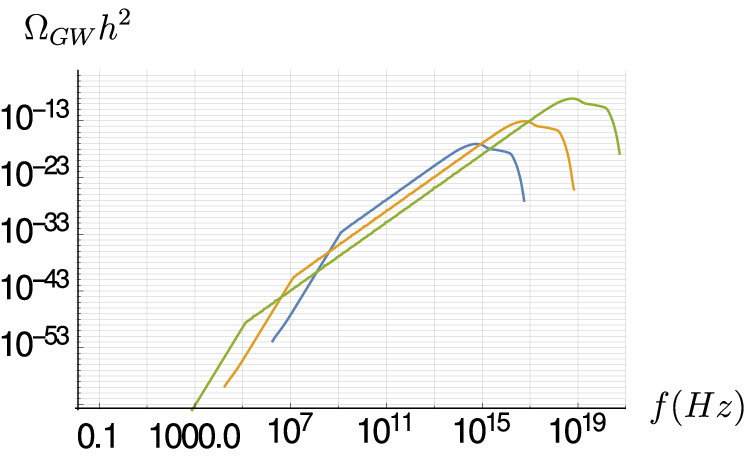}\\
    \small (b) $\beta=10^{-16}$
  \end{tabular}
  \caption{Today's gravity wave spectrum for $a_{i,*}=0.99999$. $\beta=10^{-2}$ for the upper plot, and $\beta=10^{-16}$ for the lower one. In each plot,
  blue, orange and green curves are for $\rho_i=10^{-12}, 10^{-20}, 10^{-28}$, respectively.}
\label{fig:BBN-Omega-f_rhoi}
\end{figure}

The data for ($\tau, a$) is interpolated using cubic splines to get a smooth function $a(\tau)$, in order to do the integration in Eq. (\ref{Omega-GW}).
The same discretization of $\tau$ as before is used to do 4th-order Runge-Kutta integration. The numerical error in $\Omega_{GW}$ is less than $\frac1{10}$, evaluated by
halving the $\tau$ interval and comparing the results with those shown here. Today's gravity wave spectrum is shown in Figs.
(\ref{fig:BBN-Omega-f_a}, \ref{fig:BBN-Omega-f_beta}, \ref{fig:BBN-Omega-f_rhoi}).

Fig. (\ref{fig:BBN-Omega-f_a}) shows the effects of the rotational parameter $a_{*,i}$. Within a certain range, higher $a_{*,i}$ enhances graviton emission by up to a
factor of $10^3$. This enhancement gets less significant at lower frequencies, which is consistent with the trend in Fig. (\ref{figQ-a-x}) that superradiance has smaller
effects at lower $\omega$. For high frequencies, however, higher $a_{*,i}$ produces marginally lower gravity wave signal. This is because PBHs are
effectively  Schwarzschild at late times, and the shorter PBH's lifetime for higher $a_{*,i}$ makes gravitons experience slightly higher redshifts.

Figs. (\ref{fig:BBN-Omega-f_beta}) show the effects of the initial BH fraction $\beta$. Larger PBH fraction at production generally produces larger signals. This is most significant
for not very high frequencies. $\Omega_{GW}$ is not proportional to $\beta$, because the cosmic history is also altered by the presence of PBHs. At
the high ends of frequencies, this difference diminishes gradually, because the redshifts experienced by gravitons emitted near the end of PBH's
evaporation are determined purely by the PBH's lifetime. For $\beta \ge 10^8$, PBHs dominate the universe near their end of life, so $\Omega_{GW}$'s for
different $\beta$ converge. This convergence happens earlier for lower $\rho_i$, because PBHs evolve for a larger number of e-folds, making them easier
to dominate before evaporation.

Figs. (\ref{fig:BBN-Omega-f_rhoi}) show the effects of $\rho_i$. As before, more e-folds experienced by PBHs for lower $\rho_i$ make the graviton's
frequency $f$ cover a larger range. Especially, the peak value of $\Omega_{GW}$ happens at a higher frequency. Also, as seen in
Fig. (\ref{fig:rho-a}), PBHs have larger energy density during their evaporation for larger $\rho_i$, thus producing a larger gravity wave signal
today. This effect is more apparent for $\beta=10^{-2}$ than for $\beta=10^{-16}$, also consistent with Fig. (\ref{fig:rho-a}).

The largest signal is produced in the case with $\rho_i=10^{-12}$, $\beta=10^{-2}$ and $a_{*,i}=0.99999$. The $\Omega_{GW}$ amplitude reaches approximately
$10^{-7.5}$ at $f\approx 4\times 10^{14}Hz$.

For comparison, our results agree well with \cite{Anantua:2008am} for Schwarzschild PBHs, where the greybody factors were assumed to be proportional
to $\omega^2$ without a full numerical calculation.

\section{Constraints on PBH's initial mass fraction}
\label{section:constraints}

\begin{figure}[h]
\captionsetup{justification=raggedright,
singlelinecheck=false
}
\includegraphics[width=3.4in]{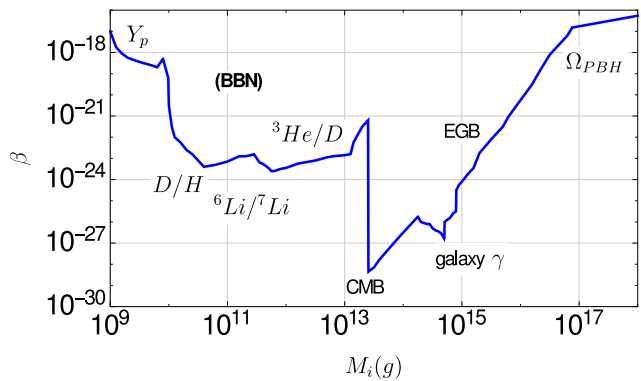}
\caption{Constraints on the PBH's mass fraction at the time of formation. The physical constraints are put below or above the corresponding segments.
BBN: from the measurements of ${^4}He$ abundance ($Y_p$), $D/H$ ratio ($D/H$), ${^6}Li/{^7}Li$ ratio (${^6}Li/{^7}Li$) and ${^3}He/D$ ratio (${^3}He/D$);
 CMB: from CMB anisotropy measurements; EGB: from extragalactic gamma ray backgrounds; galaxy $\gamma$: from gamma ray emission in the local galaxy; 
 $\Omega_{PBH}$: from the energy fraction of PBHs existing today. For details, please refer to \cite{Carr:2009jm}.}
\label{fig-constraints}
\end{figure}

While PBH which evaporate before the BBN are not seriously constrained, this is not so for PBH which evaporate later.
In this section, we constrain the abundance of such PBHs at the time of their formation.
The energy fraction of the universe in PBHs at their formation is well constrained from various observations \cite{Carr:2009jm,Green:1997sz,Ricotti:2007au,Tashiro:2008sf,Carr:2003bj,Pani:2013hpa}, especially
 for small ones which have evaporated by today. We re-plot the most recent results \cite{Carr:2009jm} in the Fig. (\ref{fig-constraints}). The dip at the critical mass $M_i\approx 5\times 10^{14}g$ corresponds
to the PBH which just completed evaporation today. The physical meaning of the various constraints is explained below, and more details can be found in \cite{Carr:2009jm}.

${^4}He$ abundance and $D/H$ ratio for $M_i=10^9-10^{10}$g: Some of the quarks and gluons emitted by PBHs fragment into mesons and antinucleons, which have enough time to get thermalized and then scatter
with background nucleons, causing interconversion between protons and neutrons. As a result, this can increase the neutron/proton ratio, and thus change the ${^4}He$ and deuterium, $D$, abundance.
The observed ${^4}He$ abundances and $D/H$ ratio thus set an upper limit on $\beta(M_i)$.

${^6}Li/{^7}He$ ratio for $M_i=10^{10}-10^{12}$g: High-energy nucleons, produced during the PBH evaporation and subsequent fragmentation, can scatter off background
nuclei from BBN. Hadrodissociation produces energetic debris including tritium, $T$, and ${^3}He$, which then scatter with the nuclei background and result in extra production of
${^6}Li$. In this way, $\beta(M_i)$ can be constrained by the well observed ${^6}Li/{{^7}Li}$ ratio.

${^3}He/D$ ratio for $M_i=10^{12}-10^{13}$g: At the temperature of PBH evaporation, the produced energetic neutrons have time to decay before hadrodissociation, emitting photons with
energies comparable to nuclei binding energy. So the photons can dissociate them and as a result, overproduce ${^3}He$ or $D$. The observation of the ${^3}He/D$ ratio thus constrains $\beta(M_i)$.

Diffuse extragalactic $\gamma$ ray background for $M_i=1.8\times 10^{14}-10^{17}$g: PBHs produce photons, both through direct Hawking radiation, and also through secondary processes involving the emitted quarks and gluons. The total $\gamma$ ray spectrum should be within
the spectrum of the diffuse extragalactic $\gamma$ ray background observed by HEAO 1, COMPTEL, EGRET and Fermi LAT. This gives un upper limit on $\beta(M_i)$.

Galaxy $\gamma$ ray background for $M_i=5\times 10^{14}-10^{15}$g: If there are PBHs residing our galactic halo, an anisotropic $\gamma$ ray background will be produced. Requiring this spectrum to be consistent with the EGRET observation, we can get the constraints on $\beta(M_i)$.

Cosmic microwave background, CMB, for $M_i=2.5\times 10^{13}-1.8\times 10^{14}$g: The PBH's emission of electrons and positrons after recombination would cause the damping of small-scale CMB anisotropies, whose measurements thus constrains $\beta(M_i)$.

PBH's relic abundance, $\Omega_{PBH}$, for $M_i\ge 10^{17}$g: For these nearly non-evaporating PBHs, this constraint comes from the requirement that today's PBH mass density is not larger than that of the cold dark matter.

Note that the above constraints are given for Schwarzschild PBHs, but in the following the same constraints will be assumed for Kerr PBHs since there 
are no constraints for rotating BHs published in the literature. We consider here PBHs with initial mass not above $10^{18}$g, above which 
the graviton signals will decrease with increasing initial masses, as will be shown.

Under these constraints, PBHs are always subdominant before evaporation. So we can use the standard cosmology model instead of solving the coupled set of equations
(\ref{complete-evolution}). The standard model parameters are taken as $h=0.673$, $\Omega_m=0.315$ and $\Omega_\Lambda=0.685$ \cite{Agashe:2014kda}.

\section{Gravity waves from PBHs with $M_i\ge 10^9 g$}
\label{section:completely}
These large PBHs have completed their evaporation after BBN, or are still evaporating today. Within the constraints on their initial mass fraction, we
calculate the largest possible gravity wave signals. For this purpose, we choose PBHs corresponding to the local maximums in Fig. (\ref{fig-constraints}),
and also $M_i=10^{15}$, $10^{16}$, $10^{17}$ and $10^{18}$g. For higher PBH masses, as shown later, the gravity wave signal is smaller.

\begin{figure}[h]
\captionsetup{justification=raggedright,
singlelinecheck=false
}
\includegraphics[width=3.4in]{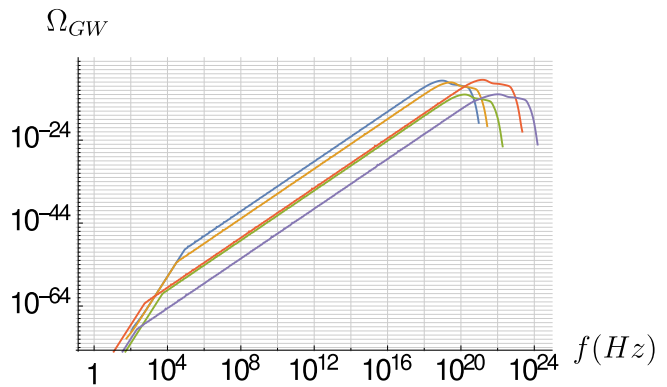}
\caption{Today's gravity wave spectrum from $M_i=10^9$g (blue), $8\times 10^9$g (orange), $2.8\times 10^{11}$g (green), $2.5\times 10^{13}$g (red)
and $1.8\times 10^{14}$g (purple). Here $a_{*,i}=0.99999$, and $\beta$'s are taken as the upper limits in Fig. (\ref{fig-constraints}).}
\label{fig:NOW-Omega-f_complete}
\end{figure}

PBHs with $M_i\lesssim 5\times 10^{14}$g have completed evaporation by today. The result for $M_i=10^9$, $8\times 10^9$, $2.8\times 10^{11}$,
$2.5\times 10^{13}$ and $1.8\times 10^{14}$g is shown in Fig. (\ref{fig:NOW-Omega-f_complete}). We only show the result for the case
$a_{*,i}=0.99999$. The difference with the $a_{*,i}=0$ case is similar to that for GUT-scale PBHs in Fig. (\ref{fig:BBN-Omega-f_a}), so it is not
shown here.

\begin{figure}[h]
\captionsetup{justification=raggedright,
singlelinecheck=false
}
\includegraphics[width=3.4in]{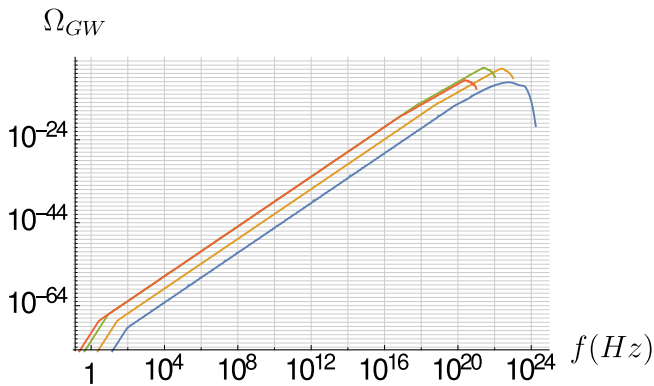}
\caption{Today's gravity wave spectrum from $M_i=10^{15}$g (blue), $10^{16}$g (orange), $10^{17}$g (green), $\times 10^{18}$g (red). Here
 $a_{*,i}=0.99999$, and $\beta$'s are taken as the upper limits in Fig. (\ref{fig-constraints}).}
\label{fig:NOW-Omega-f_exist}
\end{figure}

PBHs with $M_i\gtrsim 5\times 10^{14}$g still exist today. Using the constraints in Fig. (\ref{fig-constraints}), we get the spectrum in Fig.
(\ref{fig:NOW-Omega-f_exist}), for $M_i=10^{15}$, $10^{16}$, $10^{17}$ and $10^{18}$g. $\Omega_{GW}$ for $M_i=10^{16}$g is higher that for $M_i=10^{15}$g,
because the latter is more strictly constrained, as seen from Fig. (\ref{fig-constraints}). With the same kind of constraints, the
$M_i=10^{18}$g PBHs produce smaller signals than $M_i=10^{17}$ PBHs, because they have evaporated for a shorter time. For higher-mass PBHs,
the calculated gravity wave signal is even smaller.

\begin{figure}[h]
\captionsetup{justification=raggedright,
singlelinecheck=false
}
\includegraphics[width=3.4in]{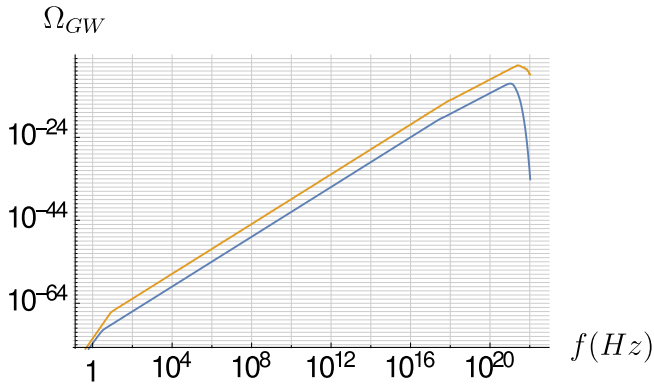}
\caption{Today's gravity wave spectrum from $M_i=10^{17}$g PBHs, with $a_{*,i}=0$ (blue) and $a_{*,i}=0.99999$ (orange). $\beta$ is taken from the
upper bound in Fig. (\ref{fig-constraints}).}
\label{fig:NOW-Omega-f_a}
\end{figure}

To see the effects of $a_{*,i}$, we plot the $M_i=10^{17}$g case also for Schwarzschild PBHs (Fig. (\ref{fig:NOW-Omega-f_a})). The enhanced emission
is clearly seen, especially in the high-frequency part. PBHs in that mass range have not shed their spin yet, so the two curves don't converge as in Fig. (\ref{fig:BBN-Omega-f_a}).

The largest amplitude of $\Omega_{GW}$ is around $10^{-6.5}$ for the case in which $M_i=10^{17}$g and $a_{*,i}=0.99999$.

\section{Conclusions}
\label{conclusions}
In this paper we addressed the question of gravity wave production from evaporating PBHs. Such PBHs can emit a good fraction of their mass into
gravitons, which can be today detected in the form of gravity waves. The spectrum of gravitational waves that can be detected today depends on
multiple factors: fraction of the total energy density which was occupied by PBHs, the epoch in which PBHs are formed, and quantities like mass and
angular momentum of evaporating PBHs.

In our analysis we used the limits on the epochs in which PBHs are formed and the total energy density occupied by PBHs at the time of their formation
from various astrophysical observations. We then calculated the greybody factors for emission of particles with various spins. These greybody factors
strongly depend on the BH angular momentum and spin of emitted particles. Highly rotating BHs dominantly emit gravitons, while non-rotating PBHs
dominantly emit scalar particles. However, angular momentum is shed faster than mass, so PBHs emit more lower spin particles toward the end of their lifetime.

The fact that
gravitons are emitted early, while the PBH is still rotating fast, means that they are redshifted more than particles emitted later in the process
of evaporation. We can see this enhanced graviton emission for higher $a_*$ in the lower frequency parts of the spectra for PBHs which have
evaporated by BBN or by today (Figs. (\ref{fig:BBN-Omega-f_a})), as well as in the whole spectrum for PBHs existing today which have not
shed their angular momentum yet. Another effect of initial angular momentum is that faster rotating PBHs, as a whole, tend to emitted more higher
energy particles, thus having a shorter lifetime. But the largest graviton signal we can observe today is from the late stage of their evolution,
because these gravitons experience smaller energy redshift. Therefore, for PBHs which have evaporated already, we see a marginally larger magnitude
in the spectrum near the maximum for initially non-rotating PBHs than the fast rotating ones (Figs. (\ref{fig:BBN-Omega-f_a})). We have examined the two
cases for $a_*=0$ and $a_*=0.99999$, for $M\omega\le 3$.

From the comprehensive analysis performed here, we conclude that very small PBHs which evaporate before BBN emit gravitons which
can give the magnitude of the gravitational wave signal today of up to $10^{-7.5}$ (Fig. (\ref{fig:BBN-Omega-f_a}, \ref{fig:BBN-Omega-f_beta},
\ref{fig:BBN-Omega-f_rhoi})). On the other hand, PBHs which are massive enough so that they are still evaporating today (and are still in agreement
with the observational constraints) can yield a signal of magnitude as high as $\sim 10^{-6.5}$ (Fig. (\ref{fig:NOW-Omega-f_exist}, \ref{fig:NOW-Omega-f_a})).
However, typical frequencies of the gravity waves from these PBHs are still too high to be observed with the current and near-future gravity wave
observations.

\vskip.2cm {\bf Acknowledgment} \vskip.2cm
This work was also partially supported by the US National Science Foundation, under Grant No. PHY-1417317.

\appendix

\section{Spin-weighted angular momentum}
\label{appendix:spin}
We will summarize ${_s}E^m_l$ for completeness. For the detailed derivation, see \cite{Fackerell:1977, Press:1973zz}. For arbitrary helicity $s$, changing the
variable from $\theta$ to $z\equiv\cos\theta$, the differential equation in question is
\be
\begin{split}
(1-z^2)\frac{d^2 S}{dz^2} & -2z\frac{dS}{dz}+\left[\gamma^2 z^2-\frac{m^2+s^2}{1-z^2} \right. \\
&\left. -\frac{2msz}{1-z^2}-2\gamma sz+ {_s}E^m_l(\gamma) \right]S=0,
\end{split}
\ee
where $\gamma=a_*M\omega$.

Define $\alpha\equiv |m+s|$, $\beta\equiv |m-s|$, we would like to express ${_s}E^m_l(\gamma)$ as a taylor series in $\gamma$,
\be
{_s}E^m_l(\gamma)=\sum^{\infty}_{p=0} {_s}f^{lm}_p \gamma^p.
\label{sElm}
\ee

\bea
{_s}f^{lm}_0 &=& l(l+1), \label{f0}\\
{_s}f^{lm}_1 &=& -2s^2m/l(l+1), \label{f1}\\
{_s}f^{lm}_2 &=& H(l+1)-H(l)-1, \label{f2}\\
{_s}f^{lm}_3 &=& 2s^2m\left[\frac{H(l)}{(l-1)l^2(l+1)}-\frac{H(l+1)}{l(l+1)^2(l+2)} \right],\label{f3}\nonumber\\
\eea
\small
\bea
{_s}f^{lm}_4 &=& 4s^4m^2\left[\frac{H(l+1)}{l^2(l+1)^4(l+2)^2}-\frac{H(l)}{(l-1)^2l^4(l+1)^2} \right] \nonumber\\
             &+& \frac12\left[\frac{H^2(l+1)}{(l+1)}+\frac{H(l+1)H(l)}{l(l+1)}-\frac{H^2(l)}{l} \right] \nonumber\\
             &+& \frac14\left[\frac{(l-1)H(l-1)H(l)}{(l-1/2)l}-\frac{(l+2)H(l+1)H(l+2)}{(l+1)(l+3/2)} \right],\nonumber\label{f4}\\
\eea
\bea
{_s}f^{lm}_5 &=& 8s^6m^3\left[\frac{H(l)}{(l-1)^3l^6(l+1)^3}-\frac{H(l+1)}{l^3(l+1)^6(l+2)^3} \right]\nonumber\\
             &+& s^2m\left[\frac{3H^2(l)}{(l-1)l^3(l+1)}-\frac{(7l^2+7l+4)H(l)H(l+1)}{(l-1)l^3(l+1)^3(l+2)} \right.\nonumber\\
             &-& \frac{3H^2(l+1)}{l(l+1)^3(l+2)}+\frac12\left(\frac{(3l+7)H(l+1)H(l+2)}{l(l+1)^3(l+3/2)(l+3)} \right.\nonumber\\
             &-& \left.\left.\frac{(3l-4)H(l-1)H(l)}{(l-2)(l-1/2)l^3(l+1)} \right)\right],\label{f5}
\eea
\bea
{_s}f^{lm}_6 &=& 16s^8m^4\left[\frac{H(l+1)}{l^4(l+1)^8(l+2)^4}-\frac{H(l)}{(l-1)^4l^8(l+1)^4} \right]\nonumber\\
             &+& 4s^4m^2\left[\frac{3H^2(l+1)}{l^2(l+1)^5(l+2)^2}-\frac{3H^2(l)}{(l-1)^2l^5(l+1)^2} \right.\nonumber\\
             &+& \frac{(11l^4+22l^3+31l^2+20l+6)H(l)H(l+1)}{(l-1)^2l^5(l+1)^5(l+2)^2}\nonumber\\
             &+& \frac12\left(\frac{(3l^2-8l+6)H(l-1)H(l)}{(l-2)^2(l-1)(l-1/2)l^5(l+1)^2} \right.\nonumber\\
             &-& \left.\left.\frac{(3l^2+14l+17)H(l+1)H(l+2)}{l^2(l+1)^5(l+3/2)(l+2)(l+3)^2}\right)\right]\nonumber \\
             &+& \frac14\left[\frac{2H^3(l+1)}{(l+1)^2}+\frac{(2l^2+4l+3)H^2(l)H(l+1)}{l^2(l+1)^2} \right.\nonumber \\
             &-& \frac{(2l^2+1)H^2(l+1)H(l)}{l^2(l+1)^2}-\frac{2H^3(l)}{l^2}\nonumber \\
             &+& \frac{(l+2)(3l^2+2l-3)H(l)H(l+1)H(l+2)}{4l(l+1)^2(l+3/2)^2}\nonumber \\
             &-& \frac{(l-1)(3l^2+4l-2)H(l+1)H(l)H(l-1)}{4(l-1/2)^2l^2(l+1)}\nonumber \\
             &+& \frac{(l+2)^2H^2(l+2)H(l+1)}{4(l+1)^2(l+3/2)^2}-\frac{(l-1)^2H^2(l-1)H(l)}{4(l-1/2)^2l^2}\nonumber \\
             &+& \frac{(l-1)(7l-3)H(l-1)H^2(l)}{4(l-1/2)^2l^2}\nonumber \\
             &-& \frac{(l+2)(7l+10)H^2(l+1)H(l+2)}{4(l+1)^2(l+3/2)^2}\nonumber \\
             &+& \frac{(l+3)H(l+1)H(l+2)H(l+3)}{12(l+1)(l+3/2)^2}\nonumber \\
             &-& \left.\frac{(l-2)H(l-2)H(l-1)H(l)}{12(l-1/2)^2l}\right]. \label{f6}\\
...... \nonumber
\eea
\normalsize
Here $H(l)=\dfrac{[l^2-(\alpha+\beta)^2/4][l^2-s^2][l^2-(\alpha-\beta)^2/4]}{2(l-1/2)l^3(l+1/2)}$.\\

The above polynomial expansion to the 6th order is accurate within $\frac1{10^4}$ for $\gamma$ up to 3.

\section{$f$ and $g$ factors}
\label{appendix:f-g}
\begin{table*}
\centering
\captionsetup{justification=raggedright,
singlelinecheck=false
}
{\footnotesize
\def\tabularxcolumn#1{m{#1}}
\begin{tabularx}{\linewidth}{@{}cXX@{}}
\begin{tabular}{c*{8}{c}}
$a_*$   & $f_0$ & $f_{1/2}$ & $f_1$ & $f_2$ & $g_0$  & $g_{1/2}$ & $g_1$ & $g_2$ \\
\hline
0.01000 & $7.429\times 10^{-5}$ & $8.185\times 10^{-5}$ & $3.366\times 10^{-5}$ & $3.845\times 10^{-6}$ & $8.867\times 10^{-5}$ & $6.161\times 10^{-4}$ & $4.795\times 10^{-4}$ & $1.064\times 10^{-4}$\\
0.10000 & $7.442\times 10^{-5}$ & $8.343\times 10^{-5}$ & $3.580\times 10^{-5}$ & $4.684\times 10^{-6}$ & $9.085\times 10^{-5}$ & $6.174\times 10^{-4}$ & $4.895\times 10^{-4}$ & $1.167\times 10^{-4}$\\
0.20000 & $7.319\times 10^{-5}$ & $8.830\times 10^{-5}$ & $4.265\times 10^{-5}$ & $7.732\times 10^{-6}$ & $9.391\times 10^{-5}$ & $6.218\times 10^{-4}$ & $5.207\times 10^{-4}$ & $1.514\times 10^{-4}$\\
0.30000 & $7.265\times 10^{-5}$ & $9.669\times 10^{-5}$ & $5.525\times 10^{-5}$ & $1.494\times 10^{-5}$ & $1.024\times 10^{-4}$ & $6.299\times 10^{-4}$ & $5.759\times 10^{-4}$ & $2.233\times 10^{-4}$\\
0.40000 & $7.097\times 10^{-5}$ & $1.089\times 10^{-4}$ & $7.570\times 10^{-5}$ & $3.116\times 10^{-5}$ & $1.125\times 10^{-4}$ & $6.430\times 10^{-4}$ & $6.599\times 10^{-4}$ & $3.603\times 10^{-4}$\\
0.50000 & $6.996\times 10^{-5}$ & $1.258\times 10^{-4}$ & $1.080\times 10^{-4}$ & $6.822\times 10^{-5}$ & $1.281\times 10^{-4}$ & $6.631\times 10^{-4}$ & $7.845\times 10^{-4}$ & $6.236\times 10^{-4}$\\
0.60000 & $7.008\times 10^{-5}$ & $1.487\times 10^{-4}$ & $1.594\times 10^{-4}$ & $1.574\times 10^{-4}$ & $1.507\times 10^{-4}$ & $6.946\times 10^{-4}$ & $9.668\times 10^{-4}$ & $1.155\times 10^{-3}$\\
0.70000 & $7.119\times 10^{-5}$ & $1.804\times 10^{-4}$ & $2.450\times 10^{-4}$ & $3.909\times 10^{-4}$ & $1.803\times 10^{-4}$ & $7.457\times 10^{-4}$ & $1.245\times 10^{-3}$ & $2.322\times 10^{-3}$\\
0.80000 & $7.969\times 10^{-5}$ & $2.284\times 10^{-4}$ & $4.014\times 10^{-4}$ & $1.104\times 10^{-3}$ & $2.306\times 10^{-4}$ & $8.366\times 10^{-4}$ & $1.706\times 10^{-3}$ & $5.286\times 10^{-3}$\\
0.90000 & $1.024\times 10^{-4}$ & $3.195\times 10^{-4}$ & $7.520\times 10^{-4}$ & $4.107\times 10^{-3}$ & $3.166\times 10^{-4}$ & $1.034\times 10^{-3}$ & $2.632\times 10^{-3}$ & $1.544\times 10^{-2}$\\
0.96000 & $1.551\times 10^{-4}$ & $4.567\times 10^{-4}$ & $1.313\times 10^{-3}$ & $1.305\times 10^{-2}$ & $4.515\times 10^{-4}$ & $1.343\times 10^{-3}$ & $3.976\times 10^{-3}$ & $4.057\times 10^{-2}$\\
0.99000 & $2.283\times 10^{-4}$ & $6.708\times 10^{-4}$ & $2.151\times 10^{-3}$ & $3.578\times 10^{-2}$ & $6.160\times 10^{-4}$ & $1.810\times 10^{-3}$ & $5.829\times 10^{-3}$ & $9.555\times 10^{-2}$\\
0.99900 & $2.625\times 10^{-4}$ & $9.253\times 10^{-4}$ & $3.057\times 10^{-3}$ & $7.251\times 10^{-2}$ & $6.905\times 10^{-4}$ & $2.340\times 10^{-3}$ & $7.723\times 10^{-3}$ & $1.753\times 10^{-1}$\\
0.99999 & $2.667\times 10^{-4}$ & $1.074\times 10^{-4}$ & $3.555\times 10^{-3}$ & $9.785\times 10^{-2}$ & $6.997\times 10^{-4}$ & $2.641\times 10^{-3}$ & $8.730\times 10^{-3}$ & $2.271\times 10^{-1}$\\
1.00000 & $2.667\times 10^{-4}$ & $1.093\times 10^{-4}$ & $3.616\times 10^{-3}$ & $1.012\times 10^{-1}$ & $7.006\times 10^{-4}$ & $2.678\times 10^{-3}$ & $8.851\times 10^{-3}$ & $2.338\times 10^{-1}$\\
\end{tabular}
\end{tabularx}
\caption{Contributions to $f$ and $g$ from each particle species. The first 14 lines are calculated, while the last one is from the cubic spline extrapolations. Please refer to the original papers
\cite{Page:1976ki, Taylor:1998dk} for the calculations.}
\label{tabf-g}
}
\end{table*}

\end{document}